\documentclass[doublecol]{epl2} 

\title{Energy landscape description of the clustering transition for active soft spheres}

\author{Moumita Maiti\inst{1} \and Michael Schmiedeberg\inst{2}}
\shortauthor{M. Maiti \etal}

\institute{                    
  \inst{1} Institut f\"ur Physikalische Chemie, Westf\"alische Wilhelms-Universit\"at (WWU), Corrensstr. 28/30, 48149 Münster, Germany\\
  \inst{2} Institut f\"ur Theoretische Physik I, Friedrich-Alexander Universit\"at Erlangen-N\"urnberg (FAU), Staudtstra{\ss}e 7, 91058 Erlangen, Germany
}


\abstract{
  For a system consisting of active soft spheres in three dimensions, we study the transition from a fluid where overlaps between particles can only occur for a short time after a collision to a state where clusters of overlapping particles persist for a long time. In order to determine the properties of the transition, we explore the energy landscape of the system in a similar way as it is done for the determination of the athermal or thermal jamming transition. Note that for zero temperature the competition of particles that attach to existing clusters and particles that detach due to thermal effects does not arise. Therefore, here we do not study such a competition because we consider systems at small or zero energy. Instead, we explore at which packing fractions and what activities cluster formation can occur at all. In case of an athermal system the transition between systems where no clusters develop at all and systems where stable clusters are found is a first order transition for packing fractions below 0.55 while the transition is continuous in case of larger packing fractions. In case of thermal systems the transition is continuous everywhere. While our approach does not deal with the real dynamics of the system, it reveals the nature of the clustering transition and it enables a deeper insight in the consequences of thermal fluctuations and the relation of the clustering transition to jamming. Though Brownian timescales diverge in athermal systems, the activity that we consider can be compared to the active velocity in other simulations if the later is measured in units of the particle size divided by an elastic time scale.}

\begin{document}

\maketitle

This is the version of the article as submitted by an author to EPL. The Version of Record is available online at https://doi.org/10.1209/0295-5075/126/46002.\\
Citation of the final version: M. Maiti and M. Schmiedeberg 2019 EPL {\bf 126} 46002\\

\section{Introduction}

Active matter, i.e., systems whose constituents can actively move, has gained a lot of attention during the last years (for reviews, e.g., see \cite{vicsek,Aranson,Marchetti,Bechinger16}). One phenomena that occurs in active particulate systems is the formation of clusters that can be observed in case of large densities and sufficiently large activity \cite{Buttinoni,Palacci,Speck,Bialke}. While in passive systems the jamming of particles at large densities occurs as soon as clusters of overlapping particles percolate, in active systems in case of sufficiently large activity clusters of overlapping particles can persist for long times at much lower densities and even without percolation. Note that often the clustering transition is discussed in terms of a competition of activity that leads to more particles ending up at a cluster and rotational diffusion that enables particles to rotate away and leave a cluster (see, e.g., \cite{Buttinoni,Palacci}). In this work we are interested at which packing fractions and what activities a cluster can form at all, because even in the limit of small or zero temperature cluster formation might be impossible at too small densities or too small activities.

Since the clustering transition can be characterized by the question whether overlaps between particles remain in the system or can be completely removed, the way of analyzing the occurring states is similar to what is used in order to study athermal jamming transition \cite{OhernLangerLiuNagel,OhernSilbertLiuNagel,liu-review}. Therefore, in this paper, we want to extend the method that has been employed to investigate the athermal jamming transition \cite{OhernLangerLiuNagel,OhernSilbertLiuNagel,Koeze18} and that was also used to explore percolation transitions related to the glass transition \cite{maiti2018,corwin,maiti2018b,maiti2018c}.

To be specific, in the approach introduced by O'Hern et al. \cite{OhernLangerLiuNagel,OhernSilbertLiuNagel} that determines how athermal jamming of soft repulsive spheres is given by the properties of the energy landscape, one starts with a random configuration and tries to remove all overlaps by minimizing the energy without crossing energy barriers. At low densities, where all overlaps can be removed, the system is called unjammed while at large packing fractions, where one cannot get rid of all overlaps by such an minimization process, the system is called jammed \cite{OhernLangerLiuNagel,OhernSilbertLiuNagel}. Recently, the athermal jamming of weakly and strongly attractive systems has been studied with a similar minimization protocol \cite{Koeze18}. While the percolation of stable clusters has turned out to be a second order transition, in weakly attractive systems a first order transition has been observed though the first order behavior might be due to finite size effects \cite{Koeze18}.

The exploration of the energy landscape can be extended to thermal systems by allowing the rare crossing of energy barriers \cite{maiti2018,maiti2018c}. As a result, one observes a thermal jamming transition that occurs at a much lower packing fraction than athermal jamming though in smaller systems or with different starting configurations the packing fraction of thermal jamming can be close to that of athermal jamming \cite{maiti2018,maiti2018c}. Thermal jamming is related to a spatial percolation transition \cite{corwin,maiti2018} and it is continuous with the same critical behavior as a directed percolation transition in time \cite{maiti2018,maiti2018c} or a random organization transition \cite{Milz2013}. In contrast, in case of the athermal jamming transition the number of contacts per particle discontinuously jumps from zero to the number required for isostaticity \cite{OhernLangerLiuNagel,OhernSilbertLiuNagel}.

In this letter we extend the approach of O'Hern et al. and study the clustering of active particles without employing dynamical simulations but by exploring the energy landscape. As we will show, unlike in a passive system the approach not necessarily leads to a completely jammed system. Overlaps can even remain in an active system that is not percolated and not isostatic. Therefore the transition that is obtained by determining whether overlaps can be removed or whether they remain despite energy minimization is a transition between a completely unjammed system and a system that consists of clusters of locally jammed particles.

Note that we mainly focus on the transition in athermal systems where there are no thermal fluctuations, no crossing of energy barriers, and where the direction of the velocity of the active particles remains constant, i.e., there is no rotational diffusion. This athermal approach can describe active systems at low temperatures or sufficiently large or even macroscopic particles, e.g. robot swarms \cite{robots}, where Brownian fluctuations can be neglected. Such athermal active systems previously have been considered, e.g., in \cite{reichhardt14}, where the transition from a collision free state in the long time limit to a state with collisions has been studied, or in \cite{liao18}, where the jamming transition is explored. The activity that we consider can be compared to the active velocity in units of the particle diameter divided by the elastic timescale as considered in conventional simulations (see, e.g., \cite{henkes14}). Since the rotational diffusion constant is zero in athermal systems, the persistence length, given by the active velocity in units of the particle diameter divided by the rotational diffusion timescale, is infinite. Similarly, the angular P{\'e}clet number, given as ratio of the elastic timescale and the rotational diffusion timescale, is infinite, too. At the end of the letter we present and discuss results for a system where the crossing of energy barriers is possible in a similar way as implemented for thermal passive particles in \cite{maiti2018,maiti2018c} and that can describe the glassy dynamics in thermal soft sphere systems \cite{maiti2018b}.

\section{Method}

We consider a system of monodisperse spheres in three dimensions that interact according to harmonic repulsive interactions, i.e., the pair interaction potential is $V(R)=\epsilon\left(r-\sigma\right)^2$, where $r$ is the distance between two spheres, $\sigma$ is the diameter of the spheres, and $\epsilon$ gives the energy scale.

Instead of simulating the real dynamics, we explore the energy landscape in the same way as in \cite{OhernLangerLiuNagel,OhernSilbertLiuNagel,maiti2018}. We start with spheres that are randomly placed with a given packing fraction $\phi=\pi\rho\sigma^3/6$, where $\rho$ is the number density, into a simulation box with periodic boundary conditions. Then the energy is minimized by employing a conjugate gradient method. The final state is characterized according to whether all overlaps can be removed or not. We assume that not all overlaps can be removed if the number of overlapping particles reaches a plateau value as a function of minimization steps (cf. \cite{OhernLangerLiuNagel,OhernSilbertLiuNagel,maiti2018,maiti2018c}). Note that in contrast to \cite{reichhardt14}, we do not only study whether collisions - corresponding to transient clusters - occur or not, but we are interested whether clusters as indicated by overlapping particles are permanently stable in the limit of long times. 

In athermal systems the minimization takes place without crossing any energy barriers. In order to mimic thermal systems, we use steps where energy barriers can be crossed with a small probability $p$ as introduced and studied for passive systems in \cite{maiti2018}. To be specific, in each step a particle is selected with a small probability $p$. A selected particle is displaced in a random direction until it reaches the nearest local minimum or maximum in that direction. It than is placed slightly behind that extremum, such that in case it was displaced upwards in energy it has crossed an energy barrier. Note that in this letter, we usually consider thermal systems except for the last section where the influence of a small but nonzero probability $p$ to cross barriers is studied.

In order to introduce an activity, at the beginning of the simulation all particles are assigned a random direction and after each minimization step of the simulation the particles are displaced by a small distance $a\sigma$ in that direction, where in the following we will call $a$ the activity of the system. Note, this is a simple model system that contains the essential ingredients of an active system in order to study the qualitative properties of the phase transition. It is not our intention to quantitatively mimic the time evolution, the forces, or the velocity due to the activity. Furthermore, the employed model can be modified in various details. In case of thermal systems, we have checked that the same results concerning the observed transition are obtained for various modifications of the protocol \cite{maiti2018}, e.g., instead of the conjugate gradient algorithm a steepest descent method can be used for minimization or the energy crossing step can be implemented in another way. Here we chose the protocol such that the energy landscape can be explored in an efficient way. If not stated otherwise, systems with $10^4$ particles are used for athermal runs and thermal runs use $10^5$ particles.

\section{State diagram of athermal systems}

 \begin{figure}[htb]
\onefigure[width=0.98\columnwidth]{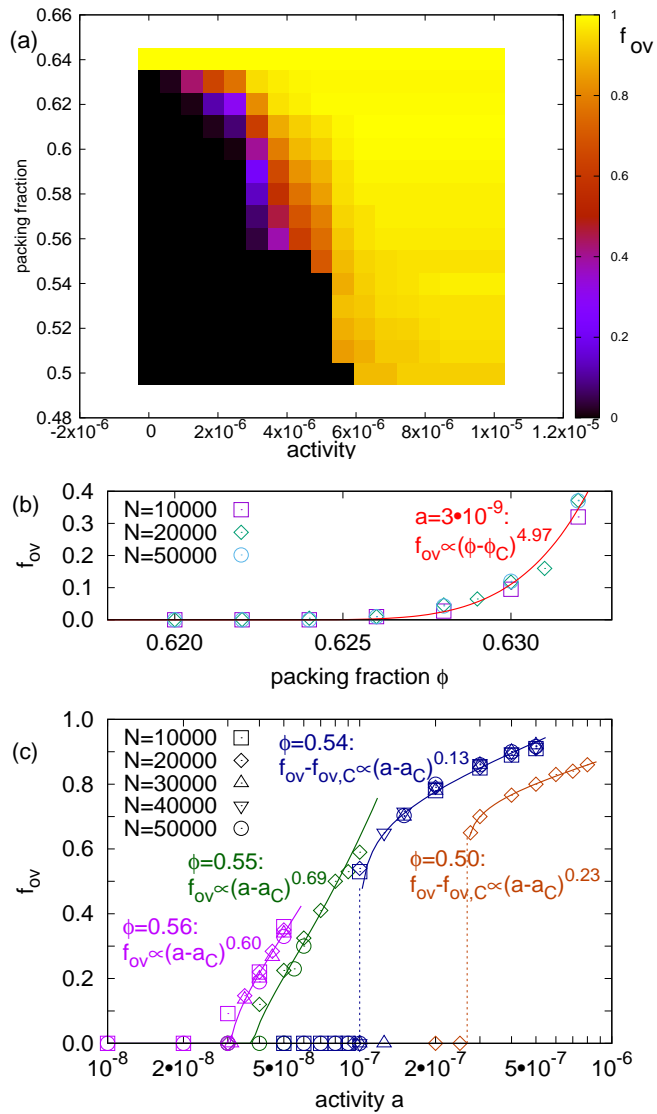}
\caption{(a) Overview of the state diagram as a function of packing fraction $\phi$ and activity $a$. The fraction $f_{ov}$ of remaining overlapping particles is denoted by the color as indicated by the color bar. Black marks final configurations without overlaps. (b,c) Fraction of overlapping particles for selected paths in the state diagram. (b) Fraction of overlapping particles for constant activity $a=3\cdot 10^{-9}$ and as a function of packing fraction $\phi$. (c) Fraction of overlapping particles as a function of activity $a$ at constant packing fractions $\phi=0.56$ (magenta), $\phi=0.55$ (green), $\phi=0.54$ (blue), and $\phi=0.50$ (orange). Different system sizes are shown. The number of particles is indicated by the symbols. While for $\phi\geq 0.55$ a continuous transition is observed, $f_{ov}$ is discontinuous for $\phi\leq 0.54$. The lines are fitted power laws.}
\label{fig.1}
\end{figure}

In fig.~\ref{fig.1}(a) the fraction of particles with overlaps that persist in the limit of infinite times (given as the limit of an infinite number of minimization steps) are shown with different colors as a function of the packing fraction $\phi$ and the activity $a$. The black color denotes systems where all overlaps can be removed such that the resulting state is a fluid. Note that the simulations are usually stopped if at some time there are no more overlaps. In principle due to the activity of the particles new overlaps could be created at later times. However, the formation of larger cluster would require the collision of more than two overlapping particles. If started from configurations without overlaps and new overlaps are created between two particles, such new overlaps are quickly removed before a third particle can stabilize the potential cluster. For the parameter regions depicted in other colors we observe states where clusters of overlapping particles exist in the long-time limit. Yellow marks the areas where almost all particles are part of a cluster of overlapping particles. Since we first consider only athermal systems the clusters cannot break up and if two clusters with opposite average velocity meet they usually stick together such that there is an overall coarsening process of the jammed clusters.

A state with overlapping particles that is either jammed or at least a state with clusters of overlapping particles occurs at large densities or large large activities. Note that jamming or even glassy dynamics usually is only expected at packing fractions above the athermal jamming transition or at least above the glass transition and that activity usually can be used to fluidize a jammed or glass-like state at least in thermal systems \cite{henkes,berthieractive}. However, it is known that clusters of overlapping particles can occur at large activities even if there are no clusters at low activities \cite{Speck,Bialke}. This behavior is confirmed by our exploration of the energy landscape. In the case where overlapping particles occur in the long-time limit, these particles are displaced due to the activity in each step of our protocol. However, the particles are forced to their previous position in the subsequent minimization step. Within a cluster of overlapping particles, there are no further rearrangements.

In order to reveal the type of the transition between a fluid state and a state with clusters, we check whether the the fraction of overlapping particles changes continuously at the transition or whether there is a jump. In figs.~\ref{fig.1}(b,c) the fraction of overlapping particles is shown along some typical paths of the state diagram. We find that the transition is continuous above or at a packing fraction of 0.55 while it is discontinuous below 0.55. If power laws are fitted to the number of overlapping particles as a function of activity for packing fractions 0.55 or slightly larger, i.e. for the cases where we observe a continuous transition (magenta and green data in fig.~\ref{fig.1}(c)), we obtain exponents in the range 0.6 to 0.7. However, in the discontinuous cases for packing fractions below 0.55 the further increase of the number of overlapping particles increases according to power laws with much smaller exponents below $0.25$.

We also studied whether our results depend on the system size. Therefore, the data in figs.~\ref{fig.1}(b,c) is shown for different system sizes (as indicated by different symbols). While for a packing fraction of 0.56 (magenta data points in figs.~\ref{fig.1}(c)) only a small deviation for small systems with $10^4$ particles close to the transition is observed, for a packing fraction of 0.55 (green data points) the curve above the transition seems to be steeper at the transition for $5\cdot 10^4$ particles (circles) then for $10^4$ particles (squares), such that the power law fit probably is affected by system size effects. For a packing fractions of 0.54 (blue data points), where the transition is discontinuous, usually all system sizes collapse on the same curve, except for a singular result with $3\cdot 10^4$ (upright triangle), where probably just by chance we observe an unclustered state.

\section{Relaxation process in athermal systems}

\begin{figure}[htb]
\onefigure[width=\columnwidth]{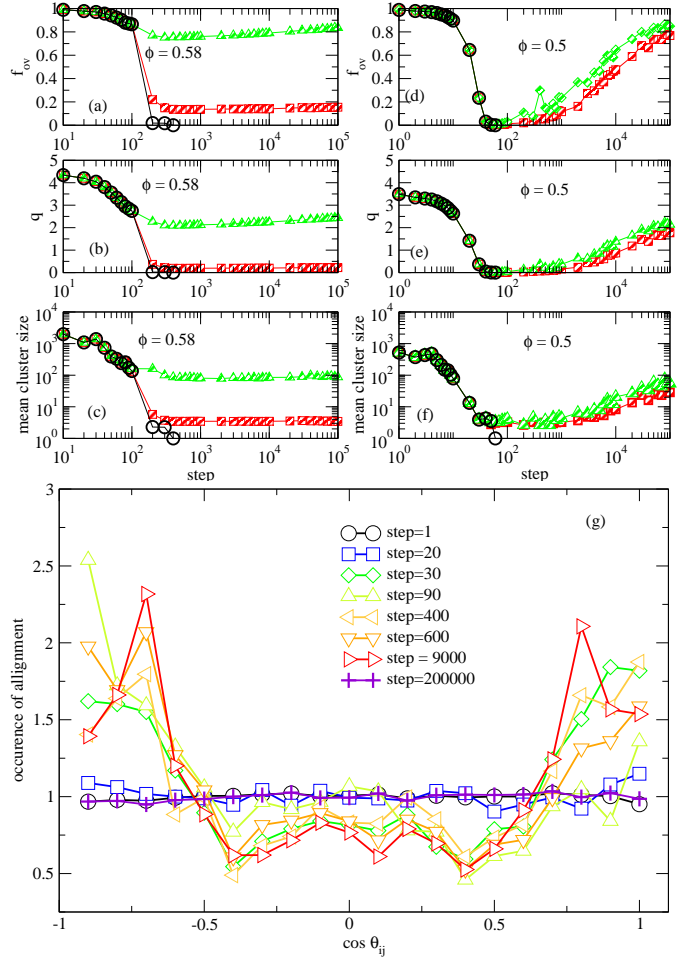}
\caption{Analysis of the relaxation process (a-c) for systems with a packing fraction $\phi=0.58$, where the transition is continuous, an (d-f,g) for systems with a packing fraction $\phi=0.5$, where the transition is discontinuous. In (a-f) always one curve in the unclustered state is shown (black), one curve that ends with clusters and is close to the transition (red), and one curve further above the clustering transition (green). To be specific, the activity for the black curves is (a-c) $a=2.5\cdot 10^{-7}$ or (d-e) $a=7.5\cdot 10^{-9}$, for the red curves (a-c) $a=5.0\cdot 10^{-7}$ or (d-e) $a=2.5\cdot 10^{-8}$, and for the green curves $a=1.0\cdot 10^{-6}$ or (d-e) $a=1.0\cdot 10^{-7}$. (a,d) Fraction of particles with overlaps, (b,e) number of contacts per particle, and (c,f) mean cluster size as a function of time. In (g) the distribution of alignments within clusters for different times is shown for $\phi=0.5$ and $a=2.5\cdot 10^{-6}$, i.e. for a system above a discontinuous clustering transition. Clusters are given by particles that overlap and single particles without overlaps are not considered to be clusters.}
\label{fig.2}
\end{figure}

For small times the fraction of particles with overlaps, the number of contacts per particle, as well as the mean cluster size dramatically decreases for systems with a packing fraction below 0.55. An example of the time evolution of these quantities is shown in the panels on the right hand side of fig.~\ref{fig.2}). While for an activity below the clustering transition (black curves) all overlaps are completely removed, for the larger activities shown in figs.~\ref{fig.2}(d-f) a few clusters of overlapping particles remain (red and green curves). At longer times additional particles join these clusters such that there is an increase of the fraction of overlapping particles as well as of the number of contacts per particle. For long times a jammed cluster is growing that slowly collects smaller clusters or the few remaining non-overlapping particles. As a result, below the packing fraction 0.55 there is a sharp jump concerning the fraction of overlapping particles in the long-time limit. In case of activities below the transition there are zero overlaps because all overlaps have been removed before the formation of a large cluster could start while the fraction of overlapping particles is almost 1 in case of activities above the transition. This is the origin of the discontinuous clustering transition below packing fraction 0.55.

For larger packing fractions jammed regions seem to occur already early in the exploration of the energy landscape. As is depicted in figs.~\ref{fig.2}(a-c) for a packing fraction of 0.58, instead of first removing almost all overlaps and then building up a cluster as it is the case below 0.55, for larger packing fractions overlaps are only removed until the steady state value is approached. While in the case without activity one can remove all the overlaps (cf. black curves) by the energy minimization as long as the packing fraction is below $\phi_J=0.64$, for larger activities some overlapping particles remain (red and green curves).

As is shown in \cite{maiti2018,corwin}, in case of packing fractions above 0.55 the system is close to a spatially percolated system, e.g., if the particles in the last step of a minimization process are kept in contact, one finds spatially percolated rigid clusters \cite{corwin}. Furthermore, even without artificially keeping the particles in contact, we discovered in \cite{maiti2018} that due to a spatial percolation above 0.55 rare rearrangements can cause large and even percolating regions of the systems to have to restart the relaxation process such that in total the non-overlapping states are never reached. In the athermal but active system that we consider here, we suspect that the activity has a similar consequence, namely that the active motion of some particles can affect a large and even percolated region of the system and as a consequence the removal of all overlaps is not possible in case the activity is large enough. Note that the transition observed in \cite{maiti2018} is continuous and that here we also observe a continuous increase of the fraction of overlapping particles that remains in the long-time limit for packing fractions above 0.55.

In fig.~\ref{fig.2}(g) the distribution of scalar products between the orientation vectors for all pairs of particles where both particles are in the same cluster are shown for different times for the case of system at packing fraction $\phi=0.5$ and activity $a=2.5\times 10^{-6}$. The distribution is divided by the initial distribution such that is shows the excess or suppression of the given relative orientations. As indicated by the peaks at a scalar product of $1$ or $-1$ in case of small times there is an excess of particles that are align in parallel or in opposite direction and that are in the same cluster (blue curve with squares in fig.~\ref{fig.2}(g)). Probably this is because initially overlapping particles that move in the same or in the opposite direction less likely loose the overlap at an early time. In the few clusters that survive or that are newly created at times between $50$ and about $10^5$ steps, there seem to be an excess of particles with angles of $3\pi/4$ or $-3\pi/4$ between their orientation vectors while there is a suppression of angles around $\pi/4$ or $-\pi/4$. In other words, in a small cluster there are two dominating groups of particles: Within one group the particles roughly points in one direction. The overlaps within the cluster then are stabilized by the other particles of the cluster that point roughly in the opposite direction. As time goes further on, there is a coarsening of clusters and in the end almost all particles are part of the same cluster such that the distribution of relative orientations looks exactly as in the beginning (see violet curve with crosses in fig.~\ref{fig.2}(g)).

\section{Thermal systems}

 \begin{figure}[htb]
\onefigure[width=0.8\columnwidth]{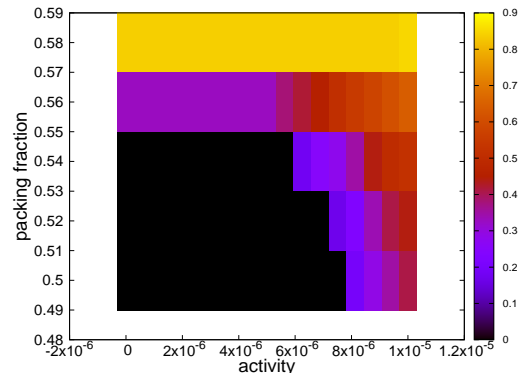}
\caption{State diagrams denoting the fraction of overlapping particles in the steady state in a thermal system where the temperature is given by the probability $p=10^{-5}$ for steps where energy barriers can be crossed.}
\label{fig.3}
\end{figure}

Finally, we want to allow the possibility to cross energy barriers. Therefore, in the same way as in \cite{maiti2018} we add random displacements of particles with a small probability $p$ that can result in the crossing of barriers. In fig.~\ref{fig.3} the resulting state diagram is shown. Interestingly even for a very small probability of $p=10^{-5}$ we find that the state diagram is different from the athermal one shown in fig.~\ref{fig.1}(a). Most importantly, overlaps cannot be completely removed for all packing fractions above 0.55 and we observe continuous transitions everywhere. Note that in case of a thermal but passive system we also have not found any state without overlaps in case of packing fractions above 0.55 and that all transitions are continuous as well \cite{maiti2018}. Therefore, even in active systems the nature of the transition from states without overlaps to states with overlaps seems to be determined by the thermal behavior. However, please note that the interpretation of the states with overlaps at large activity might be different. Here we find states consisting of clusters with overlaps that due to the rare thermal rearrangements in principle can dissolve, but since the formation of clusters occurs at the same time, one never gets rid of all clusters with overlapping particles.

\section{Discussion and Conclusions}

We show that if after each minimization step all particles are displaced according to an initially chosen direction there is a first order transition between a fluid and a cluster phase in case of an athermal system at packing fractions below $0.55$. In case of packing fractions above $0.55$ or if a thermal system is considered, the transition becomes continuous. The occurrence of a clustered phase at large densities or large activities is in agreement with results from experiments (cf., \cite{Buttinoni,Bialke}), simulations of real dynamics \cite{Palacci,Bialke} as well as mean field theories \cite{Speck,Bialke}. In case of large enough packing fractions the clustering transition probably is related to the jamming transition, namely if the locally jammed clusters that we find are percolated such that they form a globally jammed structure. Then our results can be directly compared to \cite{liao18,henkes14}, where athermal systems \cite{liao18} or at least systems \cite{henkes14} with constant rotational diffusion but varying activity are considered. Furthermore, as predicted by our approach for the thermal system, the transition upon increase of the activity seems to be continuous in case of the simulations presented in \cite{Palacci}. In contrast, the mean field description \cite{Speck} seems to allow for a jump concerning the number of overlapping particles at the transition as in our athermal case. Our approach reveals the order of the transition both in athermal and thermal systems.

We are well aware that our protocol does not correspond to real dynamics. Furthermore, the activity in our method is given by a displacement per step where the step duration can vary. However, with our approach we were able to reveal the mechanisms that lead to the different orders of the clustering transition while in case of systems with real dynamics the order and the type of the transition often is covered by the noise of the dynamics.

In case of a thermal system the rotational diffusion of the particles is neglected. The rotation of particles can lead to the dissolution of clusters in a similar way as the barrier crossing that we considered in our approach and as a consequence, we qualitatively do not expect a different behavior from the results that we obtain by allowing rare barrier crossings.

Concerning the interaction potential many modifications are possible for future investigations of the interesting phenomena and properties that have been observed in passive systems and that might be similar (or maybe different) in active systems. For example, the dynamics of soft particles in passive systems can be mapped onto hard systems \cite{BerthierandWitten,BerthierandWitten1,haxton2011,schmiedeberg2011} and multiple reentrant glass transitions are observed in ultrasoft systems \cite{berthier10,schmiedeberg13,miyazaki16}. Systems with attractive interactions are argued to be generic for jamming in nature \cite{Koeze18}. In case of even more complex attractive and repulsive contributions gel networks are observed that possess complex dynamical properties \cite{gel} and where the gelation process starts with a spatial directed percolation transition \cite{kohl}. The investigation of these properties in corresponding active systems certainly would be of great interest.

In active matter not only the clustering transition, but also complete jamming or glassy dynamics are of great interest \cite{henkes14,henkes,berthieractive}. Furthermore, in other driven non-equilibrium systems transitions that might be related are studied, e.g., in systems driven through an environment with obstacles \cite{Reichhardt18} or in sheared systems \cite{Olsson,bi,ikeda,bonn,Vagberg,Wyart,kohl17,bonn2018}. The exploration of the energy landscape in sheared systems is an important topic of recent and future research (see, e.g., \cite{heuer1,heuer2}). Our results and our approach can lead to a deeper understanding of the universal properties or the differences concerning the transition(s) that govern clustering, athermal jamming, the glass transition, gelation, yielding, clogging, and other phenomena related to the slowdown of dynamics or frustration in particulate systems.

\acknowledgments
The project was supported by the Deutsche Forschungsgemeinschaft (Grant No. Schm 2657/3-1). We gratefully acknowledge the computer resources and support provided by the Erlangen Regional Computing Center (RRZE).

\end{document}